\documentclass[
	aps,
	prb,
	reprint,
	amsmath,
	amssymb,
	groupedaddress]{revtex4-2}
	
\usepackage{amsmath}
\usepackage{amsfonts}
\usepackage{graphicx}
\usepackage{dcolumn}

\usepackage{bm}
\usepackage{braket}
\usepackage{color}
\usepackage[normalem]{ulem}
\usepackage[breaklinks=true,colorlinks]{hyperref}
\hypersetup{
    colorlinks=true,       
    linkcolor=cyan,          
    citecolor=magenta,        
    filecolor=magenta,      
    urlcolor=blue,           
    runcolor= blue
}

\begin{document}

\title{Simulated quantum annealing as a simulator of non-equilibrium quantum dynamics}

\author{Yuki Bando}
\thanks{Present address: Arithmer Inc., R\&D Headquarters, Terashimahonchonishi, Tokushima-shi, Tokushima 770-0831, Japan}\affiliation{Institute of Innovative Research, Tokyo Institute of Technology, Yokohama, Kanagawa 226-8503, Japan}

\author{Hidetoshi Nishimori}
\affiliation{Institute of Innovative Research, Tokyo Institute of Technology, Yokohama, Kanagawa 226-8503, Japan \\
Graduate School of Information Sciences, Tohoku University, Sendai, Miyagi 980-8579, Japan \\
RIKEN Interdisciplinary Theoretical and Mathematical Sciences (iTHEMS), Wako, Saitama 351-0198, Japan}

\date{\today}

\begin{abstract}
Simulated quantum annealing  based on the path-integral Monte Carlo is one of the most common tools to simulate quantum annealing on classical hardware.  Nevertheless, it is in principle highly non-trivial whether or not this classical algorithm can correctly reproduce the quantum dynamics of quantum annealing, particularly in the diabatic regime. We study this problem numerically through the generalized Kibble-Zurek mechanism of defect distribution in the simplest ferromagnetic one-dimensional transverse-field Ising model with and without coupling to the environment. 
We find that,in the absence of coupling to the environment, simulated quantum annealing correctly describes the annealing-time dependence of the average number of defects, but a detailed analysis of the defect distribution shows clear deviations from the theoretical prediction.  When the system is open (coupled to the environment), the average number of defects does not follow the theoretical prediction  but is qualitatively compatible with the numerical result by the infinite time-evolving block decimation combined with the quasi-adiabatic propagator path integral, which is valid in a very short time region. The distribution of defects in the open system turns out to be not far from the theoretical prediction. It is surprising that the classical stochastic dynamics of simulated quantum annealing ostensibly reproduce some aspects of the quantum dynamics. However, a serious problem is that  it is hard to predict for which physical quantities in which system it is reliable.
Those results suggest the necessity to exert a good amount of caution in using simulated quantum annealing to study the detailed quantitative aspects of the dynamics of quantum annealing. 
\end{abstract}


\maketitle

\section{Introduction}
Quantum annealing was originally proposed as a metaheuristic to solve classical combinatorial optimization problems~\cite{Kadowaki1998, Kadowaki1998t,Brooke1999,Farhi2001,Santoro2002,Santoro2006,Das2008, Morita2008, Albash2018, Hauke2020}. Recent years have seen its further development as a quantum simulator of materials~\cite{Harris2018, King2018, King2019, Mishra:2018, Gardas2018, Weinberg2020, Bando2020, Nishimura2020, King2021, Kairys2020}. In both of these applications, it is important to check the reliability of outputs from real quantum devices against the data obtained on classical computers by numerical solutions of the Schr\"odinger equation for closed systems (isolated from the environment) and those of quantum master equations for open systems (coupled to the environment). It is, however, difficult to carry out those numerical studies on classical hardware beyond moderate sizes because the dimension of the relevant Hilbert space increases exponentially with the number of qubits. 

Classical simulations of quantum annealing using stochastic processes are often considered as an alternative tool to study the properties of quantum annealing for large systems.  In the spin-vector Monte Carlo (SVMC)~\cite{Shin2014}, one replaces spin-$\frac{1}{2}$ Pauli operators in the Hamiltonian by classical rotors of unit length and stochastically updates the system state. This method is known to reproduce some features of the outputs from the quantum device, the D-Wave quantum annealer \cite{Boixo2014,Boixo2016,Albash2021}.  Simulated quantum annealing (SQA) is another powerful classical tool that uses the path-integral Monte Carlo~\cite{Kadowaki1998t,Santoro2002,Martonak2002,Martonak2004,Battaglia2005,Stella2006,Boixo2013,Dickson2013,Katzgraber2014,Heim2015,Albash2015,Boixo2016,Isakov2016,Mbeng2019,Nishimura2020}. The latter method is in principle designed to simulate equilibrium properties of quantum systems without a sign problem~\cite{Suzuki1976}. However, in the context of quantum annealing, it is used to simulate dynamical behaviors of the transverse-field Ising model as initially attempted in Ref.~\cite{Kadowaki1998t}. 
Recent examples include Ref.~\cite{Mbeng2019}, where the dynamics of a random Ising chain with transverse field was studied by SQA, and Refs.~\cite{Albash2016,Konz2021}, where the performance of various embedding schemes was compared using SQA.

There are few a priori reasons to expect that the classical stochastic dynamics of SQA faithfully reproduces the quantum dynamics of quantum annealing, which often operates in the regime out of equilibrium or away from the adiabatic limit.  It is nevertheless known that some aspects of the non-equilibrium dynamics of quantum annealing can be understood by SQA, notably in relation to incoherent quantum tunneling \cite{Isakov2016}.
The present paper tries to shed light on this problem through the analyses of the simplest case of the non-random one-dimensional ferromagnetic Ising model in a transverse field.  In particular, we study how far SQA reproduces the predictions of the generalized Kibble-Zurek mechanism on the distribution of defects after the system is driven across a critical point at a finite rate \cite{Adolfo2018,Gomez2020}. We also compare SQA with the direct numerical method for low-dimensional quantum systems, the infinite time-evolving block decimation (i-TEBD) combined with the quasi-adiabatic propagator path integral (QUAPI) \cite{Suzuki2019,Oshiyama2020}, which is believed to produce accurate results in a very short time region.

The problem of the generalized Kibble-Zurek mechanism in one dimension was studied extensively in Ref.~\cite{Bando2020}, where it was seen that the behavior of the D-Wave device can be understood in terms of a quantum system under the influence of an environment.  The present paper has a different point of view, i.e., not to examine the behavior of the D-Wave device but to study how far SQA is useful to explain the dynamical properties of the quantum system through comparison of the SQA data with the predictions of the generalized Kibble-Zurek mechanism as well as with the data from the i-TEBD-QUAPI.

The next section describes the problem and the methods of analyses.  Section \ref{section:results} reports the results. The final section is devoted to a summary and discussion.

\section{Problem and methods}
In the present section, we describe the problem to be studied, the methods of analysis, and the physical quantities to be observed.

\subsection{Closed- and open-system Hamiltonians}
We study the non-equilibrium dynamics, in particular the properties related to the original~\cite{Kibble1976,Zurek1985} and the generalized Kibble-Zurek mechanisms \cite{Adolfo2018,Gomez2020}, of the ferromagnetic transverse-field Ising model in one dimension with a free boundary,
\begin{equation}
\label{eq:closed_H}
H(t) = -J(t)\sum_{i=1}^{L-1}\hat{\sigma}^{z}_{i}\hat{\sigma}^{z}_{i+1} - \Gamma(t)\sum_{i=1}^{L}\hat{\sigma}^{x}_{i}.
\end{equation}
Here $L$ is the number of sites (qubits), $\hat{\sigma}^{\mu}_{i}~(\mu=x, z)$ is the $\mu$ component of the Pauli matrix at site $i$, and $J(t)$ and $\Gamma(t)$ are the time-dependent coefficients for the target and driver terms, respectively. We use the linear time dependence of the coefficients for simplicity,
\begin{eqnarray}\label{eq:schedule}
J(t)&=&\frac{t}{t_{a}}, \\
\Gamma(t)&=&1-\frac{t}{t_{a}},\label{eq:scheduleG}
\end{eqnarray}
where $t_{a}$ is the annealing time and the time $t$ runs from 0 to $t_a$. This system in the ground state undergoes an equilibrium second-order phase transition at a critical point $J=\Gamma$ in the limit of infinite system size~\cite{nishimori11book}.

In addition to the above Hamiltonian representing a closed quantum system, we also study the open system described by the Hamiltonian,
\begin{align}\label{eq:open_H}
&H(t) = -J(t)\sum_{i=1}^{L-1}\hat{\sigma}^{z}_{i}\hat{\sigma}^{z}_{i+1} - \Gamma(t)\sum_{i=1}^{L}\hat{\sigma}^{x}_{i} \nonumber \\ 
&+ \sum_{i=1}^{L}\sum_{k}\Big(V_{k}(\hat{a}_{i,k}+\hat{a}^{\dagger}_{i,k})\hat{\sigma}^{z}_{i}+\omega_{i,k}\hat{a}^{\dagger}_{i,k}\hat{a}_{i,k}\Big),
\end{align}
where $V_{k}$ is the coupling constant with the $k$th bosonic mode $\hat{a}^{\dagger}_{i,k}$ and $\hat{a}_{i,k}$ representing the environment, which is assumed to have the standard Ohmic spectral density $J(\omega)=4\pi\sum_{k}{V_k}^{2}\delta(\omega-\omega_{i,k})=2\pi\alpha\omega$.
Following Ref.~\cite{Werner2005}, we assume $J(\omega)$ to
be cutoff at some $\omega_c$, beyond which $J(\omega)$ is set to 0.
The coefficient $\alpha$ describes the magnitude of dissipation caused by the coupling to the environment. The whole system of Eq. (\ref{eq:open_H}) is isolated from other degrees of freedom, kept at zero temperature, and in principle running under the unitary Schr\"odinger dynamics without the constraint of adiabaticity.  SQA is supposed to simulate the dynamical behavior of this system.

\subsection{Method}

The problem defined above has been studied extensively for many years.   Directly pertinent to the present paper are, first, the path-integral Monte Carlo simulation by Werner et al.~\cite{Werner2005}, who studied the equilibrium phase diagram and critical exponents of the open system of Eq.~(\ref{eq:open_H}), and found that critical exponents are different from those of the closed system, Eq.~(\ref{eq:closed_H}), but are independent of the coupling strength $\alpha(>0)$.  Also, Bando et al.~\cite{Bando2020} carried out experiments on the D-Wave quantum annealers and compared the data with the generalized Kibble-Zurek mechanism~\cite{Adolfo2018,Gomez2020} to conclude that the data from the devices can be understood in terms of a system under the bosonic environment of Eq.~(\ref{eq:open_H}), not of the closed system Eq.~(\ref{eq:closed_H}), and also that the classical SVMC is unable to explain the dynamical properties of the D-Wave devices.

We combine these two contributions and study the systems of Eqs. (\ref{eq:closed_H}) and (\ref{eq:open_H}) by SQA, path-integral Monte Carlo simulations with time-dependent coefficients $J(t)$ and $\Gamma (t)$, and compare the data with those from the generalized Kibble-Zurek mechanism as well as from  the more direct numerical method of the iTEBD-QUAPI \cite{Oshiyama2020,Suzuki2019}, in which the time-dependent density operator is expressed by a  matrix-product state after a Trotter decomposition and tracing out the bosonic degrees of freedom. Then a controlled truncation of basis states allows one to perform evaluation of physical quantities.  See Refs.~\cite{Oshiyama2020,Suzuki2019} for details.

To perform SQA, we first apply the Suzuki-Trotter decomposition \cite{Suzuki1976} to the expression of the partition function for the Hamiltonian of Eq.~(\ref{eq:open_H}). The  partition function is then expressed as the corresponding partition function of a classical Ising model in two dimensions with long-range interactions in the Trotter (imaginary-time) direction~\cite{Werner2005},
\begin{equation}\label{eq:sqa}
Z(t)=Z_{0}\sum_{\{S_i(\tau)=\pm1\}} e^{-\beta_{\rm{eff}}H_{\rm{eff}}(t,\bm{S})} ,
\end{equation}
where $Z_0$ is the partition function of free bosons and
\begin{align}
H_{\rm{eff}}(t,\bm{S})&=-J(t)\sum_{i=1}^{L-1}\sum_{\tau=1}^{P}S_{i}(\tau)S_{i+1}(\tau) \nonumber \\
&\ -\frac{\gamma(t)}{\beta_{\rm{eff}}}\sum_{i=1}^{L-1}\sum_{\tau=1}^{P}S_{i}(\tau)S_{i}(\tau+1) \label{eq:Heffective} \\
&\ -\frac{\alpha}{2\beta_{\rm{eff}}}\sum_{i=1}^{L-1}\sum_{\tau>\tau'}\left(\frac{\pi}{P}\right)^2\frac{S_{i}(\tau)S_{i}(\tau')}{\sin^{2}(\frac{\pi}{P}|\tau-\tau'|)}. \nonumber
\end{align}
Here, $\beta_{\rm{eff}}=\beta/P$ with $\beta$ the inverse temperature and $P$ the number of Trotter slices along the imaginary-time axis.  The coefficient in the second line is defined as $\gamma(t)=-\frac{1}{2}\ln{[\tanh{(\beta_{\rm{eff}}\Gamma(t))}]}$. A periodic boundary condition $S_{i}({P+1})=S_{i}({1})$ is imposed along the Trotter axis.  The closed system of Eq. (\ref{eq:closed_H}) is recovered by setting $\alpha=0$.
Zero-temperature properties of Eq.~(\ref{eq:open_H}) can be simulated with sufficiently large values $\beta$ and $P$ with a fixed finite value of the ratio  $\beta_{\rm{eff}}=\beta/P$ \cite{Kadowaki1998t,Santoro2002,Santoro2006}. We choose $\beta_{\rm{eff}}=1$ and $P=4L$ to satisfy this condition as described in more detail below.

Although cluster updates have sometimes been used in related studies~ \cite{Werner2005,Heim2015,Mbeng2019}, we use the simple single-flip Metropolis-update dynamics because, first, it is non-trivial which stochastic dynamics better simulates quantum dynamics, and second, we do not expect to encounter the problem of slow relaxation, which hampers simulations of disordered systems based on simple single-spin updates~\cite{Mbeng2019}, or we do not need to reach the equilibrium necessary for the study of critical exponents~\cite{Werner2005}. The annealing time $t_a$ is identified with the total number of Monte Carlo steps divided by the number of sites $LP$ and will be denoted by $\tau_{\rm MCS}(=t_a)$. 
Values of time-dependent coefficients $J(t)$ and $\Gamma (t)$ are updated according to Eqs. (\ref{eq:schedule}) and (\ref{eq:scheduleG}) after a unit time (a Monte Carlo update trial per spin), $t \to t+1$.

\subsection{Physical quantities}

Following Ref.~\cite{Bando2020}, we measure several quantities by SQA to be compared with predictions of the generalized Kibble-Zurek mechanism and the iTEBD-QUAPI.  All quantities are measured at the end of computation where $\Gamma(t_a)=0$.

The main quantity of interest is the number of defects (spatially mis-aligned spin pairs) $n$,
\begin{equation}\label{eq:defect}
n = \sum_{i=1}^{L-1}\sum_{\tau=1}^{P}\left[1-S_{i}(\tau)S_{i+1}(\tau)\right].
\end{equation}
Notice that the true ground state of the effective Hamiltonian of Eq.~(\ref{eq:Heffective}) is perfectly ferromagnetic with $n=0$. 

We measure the statistics of defects to be denoted by $P^{\rm{SQA}}(n)$ and compare it with the prediction of the generalized Kibble-Zurek mechanism~\cite{Adolfo2018,Gomez2020}.  According to this theory, the defect distribution is bimodal and is well approximated by the Gaussian function $Q(n)$ for large systems $L\gg 1$,
\begin{equation}\label{eq:Gaussian}
Q(n)=\frac{1}{\sqrt{2\pi \kappa_2}}\exp\left[-\frac{(n-\kappa_1)^2}{2\kappa_2}\right],
\end{equation}
where $\kappa_1\equiv\langle n \rangle$ and $\kappa_2\equiv\langle (n - \langle n \rangle)^2 \rangle$. Here angular brackets $\langle \cdots \rangle$ denote the average. We quantify the difference between $P^{\rm{SQA}}(n)$ and $Q(n)$ by the $L1$ norm,
\begin{equation}\label{eq:L1}
L1=\frac{1}{2}\sum_{n} \left| Q(n) - P^{\rm{SQA}}(n) \right| .
\end{equation}
We also follow Ref.~\cite{Bando2020} and check the proximity of $P^{\rm{SQA}}(n)$ to the Boltzmann distribution of the classical Ising part of the original Hamiltonian in Eq.~(\ref{eq:closed_H}),
\begin{align}\label{eq:Boltzman}
P^{\rm{BL}}(n)=\frac{\displaystyle{L-1 \choose n}e^{-\beta_{\rm{BL}} E(n)}}{Z} \nonumber\\
Z= \sum_{n}{L-1 \choose n} e^{-\beta_{\rm{BL}} E(n)},  
\end{align}
where $E(n)$ is the energy of the classical Ising model with $n$ defects. We optimize the effective inverse temperature $\beta_{\rm{BL}}$ in $P^{\rm{BL}}(n)$ by minimizing the $L1$ norm of Eq.~(\ref{eq:L1}) (after replacing $Q(n)$ with $P^{\rm{BL}}(n)$)
such that $P^{\rm{BL}}(n)$ approximates $P^{\rm{SQA}}(n)$ as faithfully as possible. Notice that $\beta_{\rm{BL}}$ is unrelated to any physical temperature but is a parameter to fit Eq.~(\ref{eq:Boltzman}) to the data.

The number of defects $n$ in Eq.~(\ref{eq:defect}) is essentially equivalent to the residual energy per site $E_{\rm{res}}$, the difference between the achieved energy and the true ground-state energy, for which we use the expression of Ref.~\cite{Mbeng2019},
\begin{equation}\label{eq:Eres}
E_{\rm{res}} = \frac{1}{L}\sum_{i=1}^{L-1}\left(1-\frac{1}{P}\sum_{\tau=1}^{P}S_{i}(\tau)S_{i+1}(\tau)\right).
\end{equation}
Note that $E_{\rm{res}}$ can also be regarded as the average number of defects, the average over the statistics of $P^{\rm SQA}(n)$, which was denoted by $\kappa_1$ (the first cumulant or the average) above.

The data for $E_{\rm{res}}$ are to be compared with the predictions of the Kibble-Zurek mechanism, which is described qualitatively as follows.
As the system approaches a critical point, the correlation length and relaxation time grow rapidly. When the system is not close enough to the critical point, the relaxation time is large but still shorter than the time scale of annealing, and the system has time to relax to equilibrium. As the system comes closer to the critical point, the relaxation time becomes long enough and the system has no time to relax and is effectively frozen with a non-vanishing residual energy. This picture leads to the asymptotic power law of the residual energy \cite{Dziarmaga2010},
\begin{align}
E_{\rm{res}}\propto (t_a)^{-d\nu/(1+z\nu)}.
\label{eq:KZ_res_energy}
\end{align}
Here, $d=1$ is the spatial dimension and  $z$ and $\nu$ are the dynamical and correlation-length critical exponents, respectively. 

\section{Results}
\label{section:results}
We now present our results obtained by SQA.
\subsection{Closed system}
\begin{figure}[h]
\includegraphics[width=0.8\columnwidth]{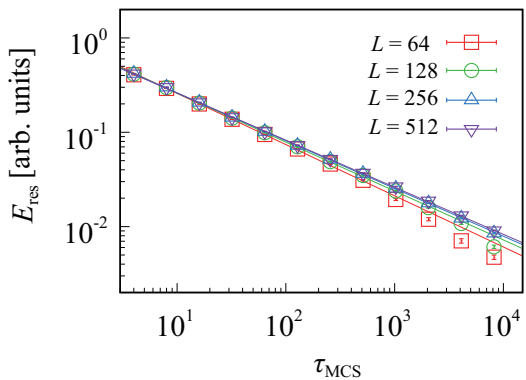}
\caption{\label{fig:nobath_KZM_scalling}Residual energy $E_{\rm{res}}$  for the closed system with sizes $L=64, 128, 256$, and $512$ as a function of the annealing time of SQA. The solid lines are fits to the power law $E_{\rm{res}}\propto(\tau_{\rm{MCS}})^{-b}$ in the range $16\le\tau_{\rm{MCS}}\le 8192$. Error bars represent the standard error of the mean computed from 100 samples, which is the case for all graphs below for the residual energy.}
\end{figure}

We first report the results for the closed system of  Eq.~(\ref{eq:closed_H}). Figure~\ref{fig:nobath_KZM_scalling} shows the residual energy $E_{\rm{res}}$ defined in Eq.~(\ref{eq:Eres}) as a function of the annealing time $\tau_{\rm{MCS}}$ for system sizes $L=64, 128, 256$, and $512$. We set $\beta_{\rm eff}=\frac{\beta}{P}=1$ and $P=4L$. 
As shown in Appendix~\ref{appendix:Trotter_size}, we have confirmed that the residual energy as a function of $P$ converges for $P=4L$. This condition is used throughout this paper.

It is clearly shown in  Fig.~\ref{fig:nobath_KZM_scalling} that the residual energy follows a power law as predicted by the original Kibble-Zurek mechanism,
\begin{align}
    E_{\rm res}\propto (\tau_{\rm MCS})^{-b}. \label{eq:power_law}
\end{align}
The values of the exponent $b$ are found as $b = 0.543\pm0.007~(L=64)$, $0.519\pm0.004~(L=128)$, $0.506\pm0.003~(L=256)$, and $0.500\pm0.003~(L=512)$. These values, particularly the last one, are in close agreement with the prediction of the Kibble-Zurek mechanism Eq.~(\ref{eq:KZ_res_energy}) for this one-dimensional closed system, which has $z=1$, $\nu=1$, and thus $b=d\nu/(1+z\nu)=\frac{1}{2}$. This agrees with previous reports by the direct numerical method of the i-TEBD-QUAPI~ \cite{Suzuki2019,Oshiyama2020} as well as by SQA~\cite{Heim2015, Mbeng2019}. By contrast, as found in Ref.~\cite{Bando2020}, the SVMC shows close, but slightly deviated, values $b=0.477\pm 0.005$ and $b=0.482\pm 0.006$, depending on the choice of the simulation temperature, 12.1 mK for the former value corresponding to the device at NASA Ames Research Center and 13.5 mK for the latter value for the device at Burnaby.

\begin{figure}[h]
\includegraphics[width=1.0\columnwidth]{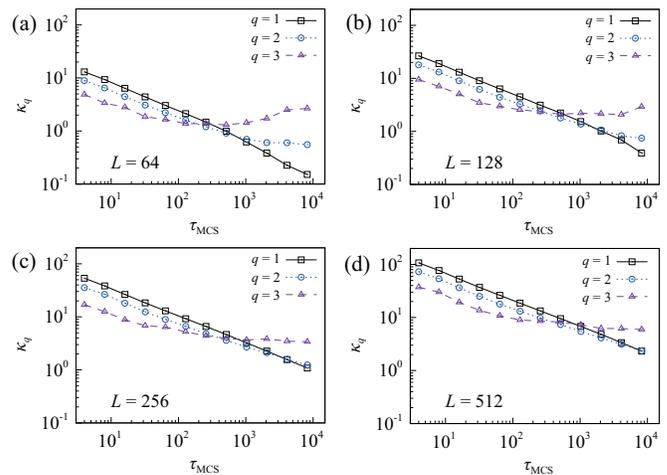}
\caption{\label{fig:nobath_cumulants}The $q$th cumulants $\kappa_q$ of defect distribution as functions of annealing time $\tau_{\rm{MCS}}$ for (a) $L=64$, (b) 128, (c) 256, and (d) 512 for the closed system.}
\end{figure}

We proceed to the analysis of the distribution function of defects $P^{\rm SQA}(n)$. Figure~\ref{fig:nobath_cumulants} shows the $q$th cumulants $\kappa_q~(q=1, 2, 3)$ of the defect distribution obtained from the final state at $\tau_{\rm MCS}$ of SQA for $L=64, 128, 256$, and $512$. 
The generalized Kibble-Zurek mechanism~\cite{Adolfo2018,Gomez2020} predicts that the second- and higher-order cumulants are proportional to the first-order cumulant (average) $\kappa_1$. The data shows that this is indeed the case for $\tau_{\rm{MCS}}$ up to about $10^2$ for any system size $L$ but the tendency changes beyond $\tau_{\rm MCS}\approx 10^2$. Comparison of data for different system sizes suggests that this deviation may possibly be a finite-size effect. 

\begin{figure}[h]
\includegraphics[width=0.8\columnwidth]{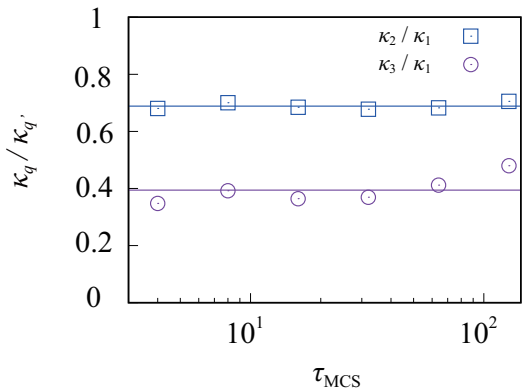}
\caption{\label{fig:nobath_cumulants_ratio}Cumulant ratios $\kappa_2/\kappa_1$ and $\kappa_3/\kappa_1$ in the range of $\tau_{\rm{MCS}}\in[4,118]$ for $L=512$ for the closed system. The constants are evaluated as $\kappa_2/\kappa_1\approx 0.688\pm0.005$ and $\kappa_3/\kappa_1\approx 0.394\pm0.019$, respectively.}
\end{figure}

Figure~\ref{fig:nobath_cumulants_ratio} shows the ratios of cumulants $\kappa_2/\kappa_1$ and $\kappa_3/\kappa_1$ for $\tau_{\rm MCS}$ up to about $10^2$ for $L=512$, which indicates the validity of proportionality of $\kappa_2$ and $\kappa_3$ to $\kappa_1$ up to $\tau_{\rm MCS}\approx 10^2$. More quantitatively, we find $\kappa_2/\kappa_1=0.688\pm0.005$ and $\kappa_3/\kappa_1=0.394\pm0.019$.  These values clearly deviate from the theoretical prediction for the closed system, $\kappa_2/\kappa_1=0.578$ and $\kappa _3/\kappa_1=0.134$~\cite{Adolfo2018,Gomez2020}. We therefore conclude that SQA does not faithfully reproduce the statistics of the defect distribution for the present closed system although the average or the residual energy is well described by SQA.

It may be useful to recall in passing that the data from the D-Wave devices showed similar proportionality but with values closer to the theory, $\kappa_2/\kappa_1=0.61\pm 0.03$ and $\kappa _3/\kappa_1=0.23\pm 0.15$ on the device at NASA and $\kappa_2/\kappa_1=0.63\pm 0.05$ and $\kappa _3/\kappa_1=0.25\pm 0.18$ on the device at Burnaby~\cite{Bando2020}. In contrast, the SVMC data marginally indicate proportionality of $\kappa_2$ to $\kappa_1$ in the short time region but with significant deviations for longer annealing times~\cite{Bando2020}. The ratio $\kappa_2/\kappa_1$ out of SVMC is close to 0.6 for the short-time region but the ratio $\kappa_3/\kappa_1$ has large error bars and it is impossible to determine its value with reliability.  

\begin{figure}[h]
\includegraphics[width=0.8\columnwidth]{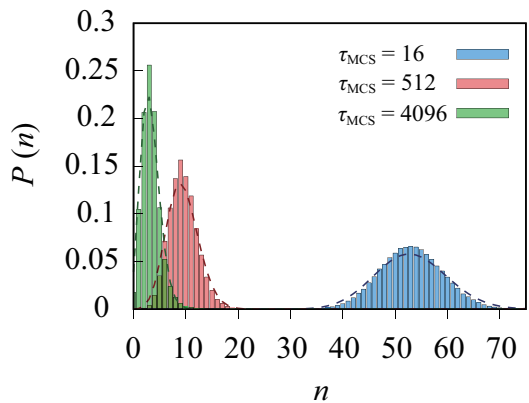}
\caption{\label{fig:nobath_defect_distribution} Probability distribution of defects by SQA at $\tau_{\rm{MCS}}=16, 512$, and $4096$ for $L=512$. The dashed lines are the Boltzmann distribution defined in Eq.~(\ref{eq:Boltzman}) with the optimized value of the effective inverse temperature $\beta_{\rm{BL}}$.}
\end{figure}

\begin{figure}[h]
\includegraphics[width=0.8\columnwidth]{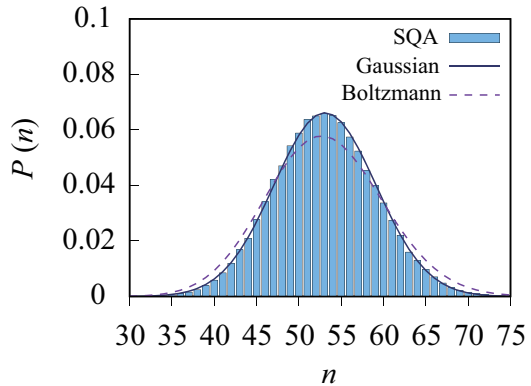}
\caption{\label{fig:nobath_gaussVSboltzman} Defect distribution at $\tau_{\rm{MCS}}=16$ for $L=512$.  The solid line is the Gaussian distribution defined in Eq.~(\ref{eq:Gaussian}), and the dashed line is the optimized Boltzmann distribution defined in Eq.~(\ref{eq:Boltzman}).}
\end{figure}

We further follow Ref.~\cite{Bando2020} and test if the defect distribution is closer to the Boltzmann or Gaussian distribution.  Figure~\ref{fig:nobath_defect_distribution} is the distribution function at $\tau_{\rm{MCS}}=16, 256$, and $4096$ and the dashed line is the Boltzmann distribution Eq.~(\ref{eq:Boltzman}) with the effective inverse temperature $\beta_{\rm{BL}}$ optimized for each $\tau_{\rm{MCS}}$.
Though the data appears to be close to the optimized Boltzmann distribution $P^{\rm{BL}}(n)$, we see a slight deviation especially at $\tau_{\rm{MCS}}=16$, which is more clearly observed in the enhanced figure in Fig.~\ref{fig:nobath_gaussVSboltzman}, where we see a better fit by the Gaussian Eq.~(\ref{eq:Gaussian}), similarly to the previous study by the D-Wave devices and SVMC~\cite{Bando2020}.

\begin{figure}[h]
\includegraphics[width=0.8\columnwidth]{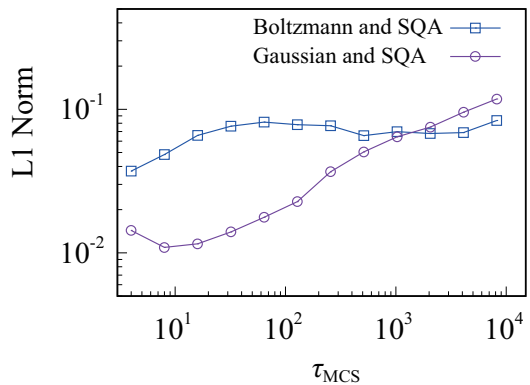}
\caption{\label{fig:nobath_L1norm} The L1 norms between the defect distribution of SQA $P^{\rm{SQA}}(n)$ and the optimized Boltzmann distribution $Q^{\rm{BO}}(n)$ (circle) and the normal distributions $Q^{\rm{GA}}(n)$ (square) for $L=512$}
\end{figure}

Figure~\ref{fig:nobath_L1norm} shows the L1 norm of the difference between the defect distribution of SQA, $P^{\rm{SQA}}(n)$, and the optimized Boltzmann distribution $P^{\rm BL}(n)$ in Eq.~(\ref{eq:Boltzman}) and the Gaussian distribution $Q(n)$ in Eq.~(\ref{eq:Gaussian}) as a function of the annealing time $\tau_{\rm{MCS}}$ for $L=512$. In the time range up to $\tau_{\rm{MCS}}\approx 10^2$, where the proportional relationship of cumulants is established, the L1 norm with the Gaussian is smaller, and thus SQA works as a Gaussian sampler rather than as a Boltzmann sampler, but for longer annealing times, e.g., $\tau_{\rm{MCS}}\approx 10^{3}$, two functions show similar degrees of proximity to the data.

Those results for the closed system indicate that the classical stochastic dynamics of SQA partly succeeds in reproducing the predictions on the quantum dynamics related to the original and generalized Kibble-Zurek mechanism. However, detailed quantitative analysis of the distribution function show deviations from the theory for the closed system.


\subsection{Open system}
\label{subsec:open_system}
We next discuss the results obtained for the open system defined in Eq.~(\ref{eq:open_H}) and simulated by Eq.~(\ref{eq:Heffective}).

\begin{figure}[h]
\includegraphics[width=0.8\columnwidth]{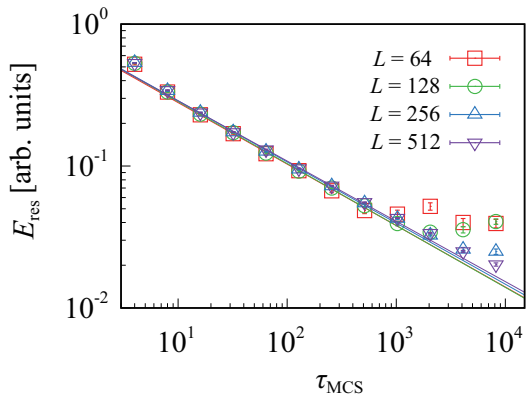}
\caption{\label{fig:bath_KZM_scalling} Residual energy $E_{\rm{res}}$ obtained by SQA for the open system with $L=64, 128, 256$, and $512$ as a function of annealing time $\tau_{\rm{MCS}}$. The solid lines are fits to $(\tau_{\rm{MCS}})^{-b}$ in the range $16\le \tau_{\rm MCS}\le 1024$ with $b\approx 0.43$ for all $L$.}
\end{figure}

Figure~\ref{fig:bath_KZM_scalling} shows the residual energy $E_{\rm{res}}$ as a function of $\tau_{\rm{MCS}}$ for $L=64, 128, 256$, and $512$ under the same parameters as before, $\beta_{\rm eff}=\frac{\beta}{P}=1$ and $P=4L$. Here, we fix the coupling strength with the boson field to $\alpha=0.6$ for all $L$.
It is seen that the power decay predicted by the Kibble-Zurek mechanism, Eq.~(\ref{eq:power_law}), holds up to $\tau_{\rm MCS}\approx 10^3$. Deviations are observed even for the largest system with $L=512$ beyond this annealing time in contrast to the case of the closed system in Fig.~\ref{fig:nobath_KZM_scalling}. It is anticipated from the behavior beyond $\tau_{\rm MCS}\approx 10^3$ that these deviations may be due to finite-size effects. Quantitatively, the exponents extracted from the linear region are $b = 0.433\pm0.013, 0.437\pm0.006, 0.431\pm0.008$, and $0.425\pm0.007$ for $L = 64, 128, 256$, and $512$, respectively, indicating deviations from the closed-system value $b=0.5$. Critical exponents $z=1.985$ and $\nu=0.638$ from equilibrium Monte Carlo simulation of the open system in Ref.~\cite{Werner2005} lead to $b=d\nu/(1+z\nu)=0.28$. Our estimate $b\approx 0.43$ from SQA with $\alpha=0.6$ is not close to either of those theoretical values 0.5 and 0.28 for closed and open systems, respectively. 

\begin{figure}[h]
\includegraphics[width=0.8\columnwidth]{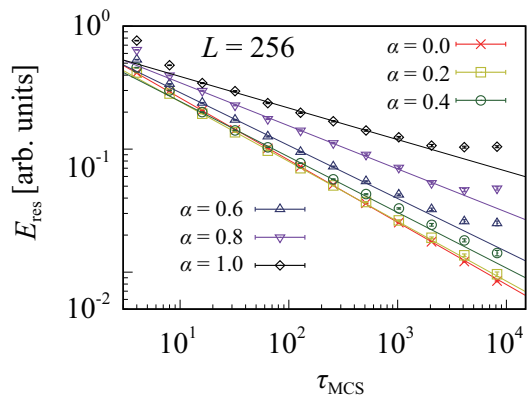}
\caption{\label{fig:bath_KZM_etas_scalling}Residual energy for different coupling constants with the environment $\alpha$ for $L=256$ as a function of annealing time $\tau_{\rm MCS}$.}
\end{figure}

Extensive numerical simulations of open quantum systems by the i-TEBD-QUAPI, which is expected to faithfully reproduce quantum dynamics for the very short-time region,  showed clear dependence of the exponent $b$ on several factors including the coupling strength $\alpha$ and the range of annealing time~\cite{Suzuki2019,Oshiyama2020,Bando2020}. We therefore checked the $\alpha$-dependence of the residual energy by SQA, and the result is in Fig.~\ref{fig:bath_KZM_etas_scalling} for $L=256$. It is seen that the exponent $b$ is a decreasing function of $\alpha$ (from $b= 0.50$ for $\alpha =0$ to $b=0.26$ for $\alpha =1.0$), consistently with the results of i-TEBD-QUAPI reported in Refs.~\cite{Bando2020,Suzuki2019,Oshiyama2020}. For the reader's convenience, we reproduce a figure from Ref.~\cite{Bando2020} as Fig.~\ref{app_fig:iTEBD} of Appendix~\ref{appendix:iTEBD} of this paper, where the defect density (proportional to the residual energy) is drawn as a function of the annealing time.  One observes there that $b$ changes from $b= 0.50$ for $\alpha =0$ to $b=0.25$ for $\alpha =1.28$, close to the values from SQA.

As suggested in Ref.~\cite{Bando2020}, taking into account the fact that the i-TEBD-QUAPI reproduces quantum dynamics for very short times, this non-universality (dependence of the exponent on $\alpha$) of the SQA data may imply that the system  is still in a transient state and much longer annealing times for much larger systems may show a value of $b$ independent $\alpha$ as expected from universality seen in the equilibrium Monte Carlo simulation \cite{Werner2005}. Viewed differently, this consistency with the quantum-mechanical simulation by the i-TEBD-QUAPI may imply that SQA, a classical stochastic process, reproduces some aspects of the non-equilibrium quantum dynamics in the short time region for the present open system.

\begin{figure}[h]
\includegraphics[width=1.0\columnwidth]{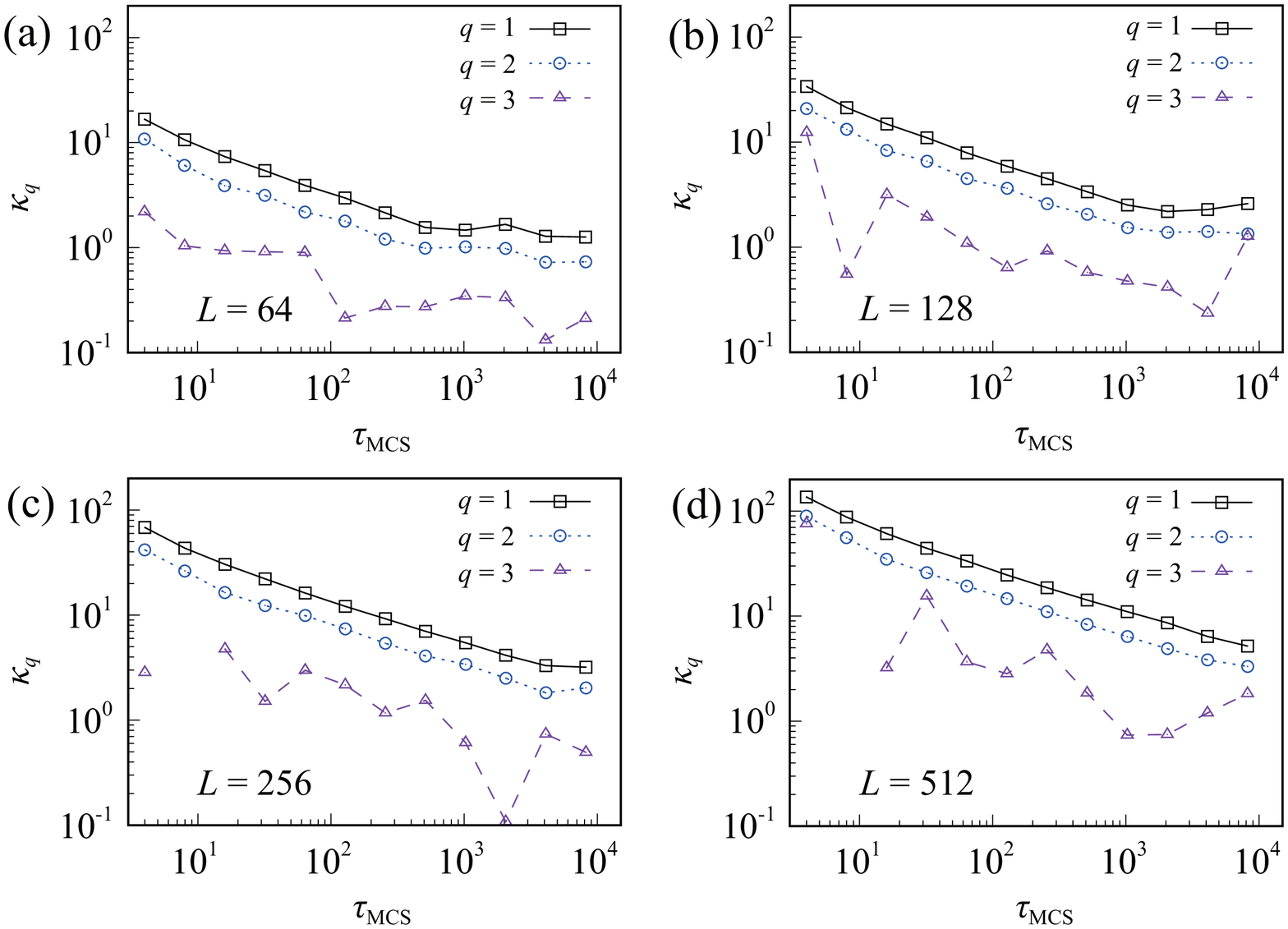}
\caption{\label{fig:bath_cumulants}Cumulants of defect distribution of open systems with coupling constant $\alpha=0.6$ for (a) $L=64$, (b) 128, (c) 256, and (d) 512.}
\end{figure}

We proceed to test if the distribution function of defects $P^{\rm SQA}(n)$ for a typical fixed value of $\alpha=0.6$ is consistent with the prediction of the generalized Kibble-Zurek mechanism as we did for the closed system. Figure~\ref{fig:bath_cumulants} shows the $q$th cumulants of the distribution as functions of the annealing time for $L=64, 128, 256$, and $512$.  It is seen that the second cumulant is proportional to the first cumulant almost up to $\tau_{\rm MCS}\approx 10^3$, a longer time range of proportionality than in the closed system depicted in Fig.~\ref{fig:nobath_cumulants}. The behavior of the third cumulant is unstable due to insufficient statistics.  

\begin{figure}[h]
\includegraphics[width=0.9\columnwidth]{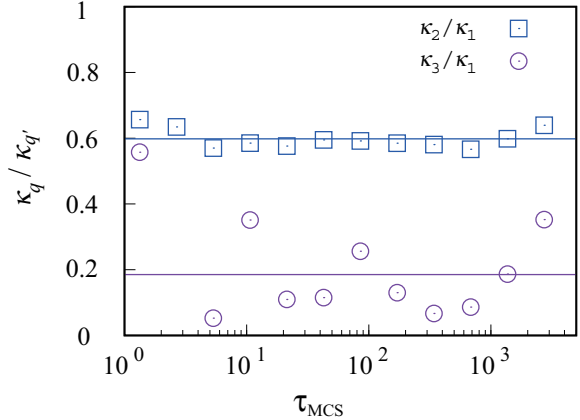}
\caption{\label{fig:bath_cumulants_ratio}Cumulant ratios $\kappa_2/\kappa_1$ and $\kappa_3/\kappa_1$ for $L=512$ for the open system with the coupling constant $\alpha=0.6$.}
\end{figure}

Figure~\ref{fig:bath_cumulants_ratio} shows the ratios of cumulants with the results $\kappa_2/\kappa_1\approx 0.598\pm0.008$ and $\kappa_3/\kappa_1\approx 0.185\pm0.048$. These values for the open system are closer to the theoretical prediction of the generalized Kibble-Zurek mechanism ($\kappa_2/\kappa_1=0.578$ and $\kappa _3/\kappa_1=0.174$~\cite{Gomez2020}) than in the case of SQA for the closed system discussed in the preceding section ($\kappa_2/\kappa_1\approx 0.688\pm0.005$ and $\kappa_3/\kappa_1\approx 0.394\pm0.019$).
We have also calculated cumulant ratios with a different coupling strength $\alpha=1.0$ in $L=512$ to see the $\alpha$ dependence. The result is $\kappa_2/\kappa_1\approx 0.569\pm0.015$ and $\kappa_3/\kappa_1\approx 0.196\pm0.135$, indicating very weak dependence on $\alpha$.

It was found in Ref.~\cite{Bando2020} that the numerical data by the i-TEBD-QUAPI for the open system indicate a similar value 0.6 for $\kappa_2/\kappa_1$ independent of $\alpha$ but it is difficult to evaluate $\kappa_3/\kappa_1$.
We have thus found that SQA for the open system successfully reproduces the prediction of the generalized Kibble-Zurek mechanism for the cumulant ratios, but the behavior of the average (residual energy) is non-universal, which is not compatible with the theory. Also, agreement with the data of the iTEBD-QUAPI in the short time region has been observed.

\begin{figure}[h]
\includegraphics[width=0.8\columnwidth]{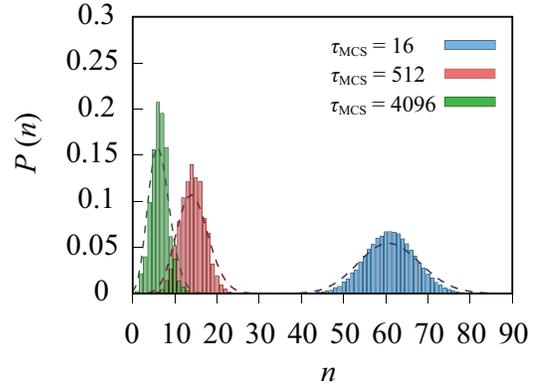}
\caption{\label{fig:bath_defect_distribution}Distribution of defects at $\tau_{\rm{MCS}}=16, 512$, and $4096$ for $L=512$ for open systems with the coupling constant $\alpha=0.6$. The dashed lines are the optimized Boltzmann distribution.}
\end{figure}

\begin{figure}[h]
\includegraphics[width=0.8\columnwidth]{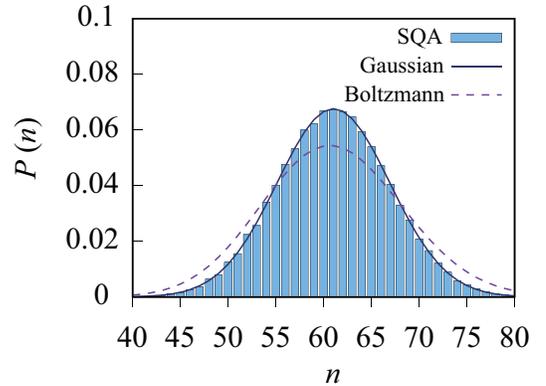}
\caption{\label{fig:bath_gaussVSboltzman} Defect distribution at $\tau_{\rm{MCS}}=16$ for $L=512$ of the open system with $\alpha=0.6$. The solid line is the Gaussian distribution in Eq.~(\ref{eq:Gaussian}) and the dashed line is the optimized Boltzmann distribution of Eq.~(\ref{eq:Boltzman}).}
\end{figure}

As in the case of the closed system, we further check whether the distribution function is closer to the Gaussian or Boltzmann. Figure~\ref{fig:bath_defect_distribution} shows the distribution function for $\tau_{\rm{MCS}}=16, 256$, and 4096 for $L=512$ with the coupling constant $\alpha=0.6$ fitted to the Boltzmann distribution Eq.~(\ref{eq:Boltzman}) with optimized effective temperature.
Although the gross feature is captured by the Boltzmann distribution, we find clear deviations. In fact, as seen in Fig.~\ref{fig:bath_gaussVSboltzman}, the Gaussian distribution better matches the data, as anticipated from the smallness of the third cumulant.

\begin{figure}[h]
\includegraphics[width=0.8\columnwidth]{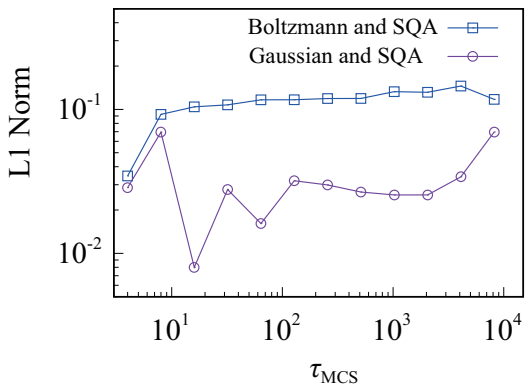}
\caption{\label{fig:bath_L1_norm} Distance, the L1 norm, between the defect distribution $P^{\rm{SQA}}(n)$ of SQA and the optimized Boltzmann distribution $P^{\rm{BL}}(n)$ (circle) and the Gaussian distribution $Q(n)$ (square) for $L=512$.}
\end{figure}

Figure~\ref{fig:bath_L1_norm} shows the L1 norm between the defect distribution of SQA, $P^{\rm{SQA}}(n)$, and the optimized Boltzmann distribution $P^{\rm BL}(n)$ in Eq.~(\ref{eq:Boltzman}) and the Gaussian distribution $Q(n)$ of Eq.~(\ref{eq:Gaussian}) as a function of the annealing time $\tau_{\rm{MCS}}$ for $L=512$. It is seen that the data is closer to the Gaussian distribution than to the Boltzmann, similarly to the case of the closed system and in agreement with the generalized Kibble-Zurek mechanism which predicts small values of higher-order cumulants than the second order, meaning Gaussian approximately. Thus SQA is closer to a Gaussian sampler rather than a Boltzmann sampler in the present open system as well.  A similar conclusion was drawn for the D-Wave device \cite{Bando2020}, and therefore we should be careful when we use quantum annealing (simulated or on the real device) for sampling purposes at least as long as the present one-dimensional system is concerned.

\section{Discussion}

We have performed numerical tests to check if simulated quantum annealing (SQA) is able to describe the non-equilibrium quantum dynamics of the simple one-dimensional ferromagnetic transverse-field Ising model across its critical point using the generalized Kibble-Zurek mechanism and the direct quantum numerics of i-TEBD-QUAPI as tools to measure the degree of success of SQA.  This is a highly non-trivial problem because the classical algorithm SQA has no a priori reason to reproduce quantum dynamics, although SQA has often been used to simulate quantum annealing, in particular for large systems.  It is nevertheless known that the phenomenon of incoherent quantum tunneling is well simulated by SQA if there exists a finite number of energy barriers between local minima \cite{Isakov2016}, a prototypical energy landscape of a simple first-order phase transition.  We have studied the problem from the point of view of non-equilibrium dynamics across a critical point (second-order transition point) in the open system keeping in mind that the data from SQA is known to agree with the prediction of the original Kibble-Zurek mechanism in the closed one-dimensional system without disorder \cite{Mbeng2019}.

We  presented detailed quantitative comparison of defect distribution with the asymptotically exact theory of the generalized Kibble-Zurek mechanism and the i-TEBD-QUAPI, the latter of which is expected to faithfully reproduce quantum dynamics in the very short time region in the present one-dimensional system.
Our results indicate that some, but not all, of the dynamical properties of the quantum system can be reproduced by SQA for the one-dimensional system:
In the absence of coupling to the environment, SQA correctly describes the annealing-time dependence of the residual energy, but the ratios of cumulants of defect distribution clearly deviate from theoretical values.  When the system is open, the residual energy does not follow the prediction of the Kibble-Zurek mechanism but shows compatibility with the numerical result by the iTEBD-QUAPI. The ratios of cumulants for the open system turn out to be closer to the theoretical values of the generalized Kibble-Zurek mechanism than in the case of the closed system. The distribution of defects is closer to the Gaussian function than to the Boltzmann distribution in both closed and open systems, which is also the case in the D-Wave quantum annealers Ref.~\cite{Bando2020}, meaning that SQA and the D-Wave device serve as Gaussian samplers, not as Boltzmann samplers.

It is remarkable that the classical stochastic dynamics of SQA ostensibly reproduces some of the properties of quantum dynamics for the present system. Nevertheless, it is hard to predict when it is reliable for what physical quantities, which is a serious problem in practical applications.
It is therefore necessary to stay very cautious when one uses SQA to clarify the dynamical properties of quantum annealing in a detailed quantitative way. It is also desirable to establish a theoretical framework to explain the present numerical observations. Though there exists a formal mapping between classical stochastic dynamics and a quantum-mechanical system \cite{Somma2007,Nishimori2015}, this is still far from sufficient to understand the results found in the present paper. It is also known that the convergence condition of SQA in the long-time limit has a very similar expression to that of  quantum annealing \cite{Morita2006,Morita2008}.  Whether or not this is a coincidence is an interesting topic closely related to the present work.

\begin{acknowledgments}
The research is based partially upon work supported by the Office of the Director of National Intelligence (ODNI), Intelligence Advanced Research Projects Activity (IARPA) and the Defense Advanced Research Projects Agency (DARPA), via the U.S. Army Research Office contract W911NF-17-C-0050. The views and conclusions contained herein are those of the authors and should not be interpreted as necessarily representing the official policies or endorsements, either expressed or implied, of the ODNI, IARPA, DARPA, or the U.S. Government. The U.S. Government is authorized to reproduce and distribute reprints for Governmental purposes notwithstanding any copyright annotation thereon.
\end{acknowledgments}

\appendix
\section{Trotter-size dependence}\label{appendix:Trotter_size}
\begin{figure}[h]
\includegraphics[width=\columnwidth]{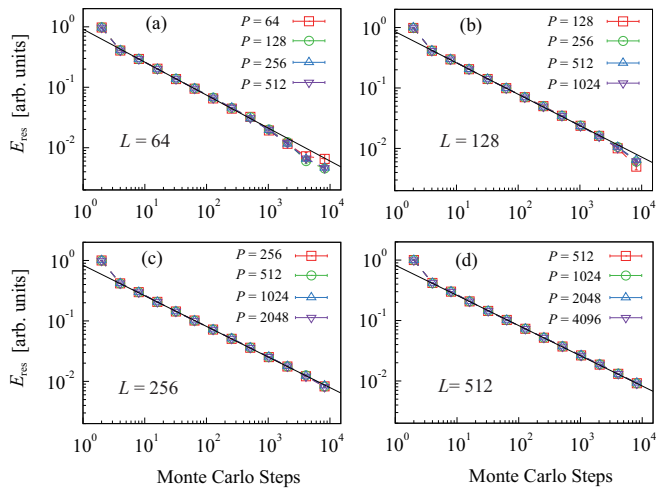}
\caption{\label{app_fig:trot_size}Residual energy $E_{\rm{res}}$ for (a) $L=64$, (b) 128, (c) 256, and (d) $512$ as a function of Monte Carlo steps or annealing time $\tau_{\rm{MCS}}$. We change the Trotter size from $P=L$ to $P=8L$.}
\end{figure}

Here, we show the Trotter-size dependence of the residual energy. Figure~\ref{app_fig:trot_size} shows the residual energy for $L=64, 128, 256$, and $512$ as a function of $\tau_{\rm{MCS}}$ for a series of Trotter size $P$. This figure shows that $P=4L$ is enough to reach convergence of the residual energy for those system sizes and annealing time.

\section{Residual energy by the i-TEBD-QUAPI}\label{appendix:iTEBD}
\begin{figure}[h]
\includegraphics[width=0.95\columnwidth]{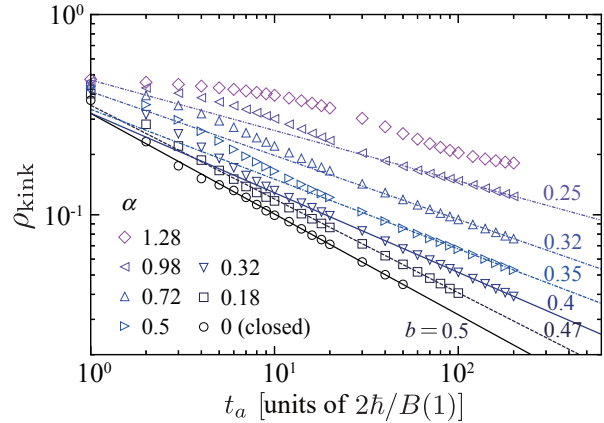}
\caption{\label{app_fig:iTEBD}
Density of defects (proportional to the residual energy) obtained by i-TEBD-QUAPI. $t_a$ is the annealing time measured in units of $\hbar$ divided by the final coupling strength $J(t=t_a)$ (denoted $B(1)/2$ in the figure) for various values of $\alpha$.  Taken from Ref.~\cite{Bando2020}.
}
\end{figure}

In Fig.~\ref{app_fig:iTEBD}, we reproduce Fig. 4(b) of Ref.~\cite{Bando2020}, where the density of defects (kinks), proportional to the residual energy, is plotted as a function of the annealing time obtained from the extensive numerical computation of i-TEBD-QUAPI. The exponent $b$ for the decay of the residual energy, Eq.~(\ref{eq:power_law}), is seen to depend on the coupling constant $\alpha$ with the bosonic environment in reasonable agreement with the SQA result described in Sec~\ref{subsec:open_system}.
\providecommand{\noopsort}[1]{}\providecommand{\singleletter}[1]{#1}%

\end{document}